\documentclass[useAMS,usenatbib]{mn2e}
\usepackage{epsfig}
\usepackage{color}
\usepackage{natbib}
\usepackage{deluxetable}
\usepackage{amssymb}
\usepackage{rotating}
\usepackage{graphicx}
\usepackage{IEEEtrantools}
\usepackage[colorlinks=true,citecolor=blue]{hyperref}
\usepackage{amsmath}
\def\apjl{{ApJ}}%
\def\apjs{{ApJS}}%
\def\aap{{A\&A}}%
\title[Weak emission line Quasars]{ 
Polarimetric and spectroscopic study of radio-quiet weak emission line quasars}
\author[Kumar et al.] {{P. Kumar$^{1,2}$\thanks{E-mail: parveen@aries.res.in (PK)},
    H. Chand$^{1}$, R. Srianand$^{3}$, C. S. Stalin$^{4}$}, P. Petitjean$^{5}$, Gopal-Krishna$^{6}$ \\
$^{1}$Aryabhatta Research
  Institute of Observational Sciences (ARIES), Manora Peak, Nainital,
  263002 India\\ $^{2}$Pt. Ravishankar Shukla University, Raipur, 492010, India\\ $^{3}$Inter-University Centre for Astronomy and Astrophysics (IUCAA), Postbag 4, Ganeshkhind, Pune 411 007, India\\$^{4}$Indian Institute of Astrophysics, Block II, Koramangala, Bangalore-560034, India\\  $^{5}$Institut d$^\prime$Astrophysique de Paris, CNRS-UPMC, UMR 7095, 98bis bd Arago, 75014 Paris, France\\
$^{6}$UM-DAE Centre for Excellence in Basic Sciences (CEBS), Mumbai 400098, India
}
\begin{document}
\date{Accepted ---. Received ---; in original form ---}

\pagerange{\pageref{firstpage}--\pageref{lastpage}} \pubyear{2018}

\maketitle

\label{firstpage}
\begin{abstract} A small subset of optically selected
  radio-quiet quasars showing weak or no emission lines may turn out
  to be the elusive radio-quiet BL Lac objects, or simply be
  radio-quiet QSOs with a still-forming/shielded broad line region
  (BLR). High polarisation ($p$ $>$ 3$-$4$\%$), a hallmark of BL Lacs,
  can be used to test whether some optically selected `radio-quiet
  weak emission line quasars' (RQWLQs) show a fractional polarisation
  high enough to qualify as radio-quiet analogs of BL Lac objects.
  Out of the observed six RQWLQs candidates showing an insignificant
  proper motion, only two are found to have $p$ $>$ 1$\%$. For these
  two RQWLQs, namely J142505.59$+$035336.2, J154515.77+003235.2, we
  found polarisation of 1.03$\pm$0.36$\%$, 1.59$\pm$0.53$\%$
  respectively, which again is too modest to justify a (radio-quiet)
  BL Lac classification. We also present here a statistical comparison
  of the optical spectral index, for a set of 40 RQWLQs with
  redshift-luminosity matched control sample of 800 QSOs and an
  equivalent sample of 120 blazars. The spectral index distribution of
  RQWLQs is found to differ, at a high significance level, from that
  of blazars and is consistent with that of the ordinary
  QSOs. Likewise, a structure-function analysis of photometric light
  curves presented here suggests that the mechanism driving optical
  variability in RQWLQs is similar to that operating in QSOs and
  different from that of blazars. These findings are consistent with
  the common view that the central engine in RQWLQs, as a population,
  is akin to that operating in normal QSOs and the primary differences
  between them might be related to differences in the BLR.
\end{abstract}
\begin{keywords}
galaxies: galaxies: active – BL Lacertae objects: general – galaxies: jet – galaxies:polarization – techniques: emission lines – quasars: general.
\end{keywords}


\section{Introduction}
BL Lac objects are characterized by a large flux variability from
radio-to-TeV frequencies, nearly featureless optical spectrum and in
many cases superluminal motion of the nuclear radio blobs~\citep[][and
  reference therein]{Urry1995PASP..107..803U}. They also have a
compact core-dominated radio morphology and usually a high ($>$ 3$\%$)
and variable polarisation at radio-to-optical frequencies. Such
observed properties of BL Lac are well explained by a model in which
the dominant source of the observed emission is a relativistic jet of
radiating plasma directed within a small angle to our line of sight,
as envisioned in the unification scheme of active galactic nuclei
(AGN)~\citep[e.g.,][]{Begelman1984RvMP...56..255B,Urry1995PASP..107..803U,Antonucci2012A&AT...27..557A}.\par
Traditionally, BL Lac objects have been discovered from radio and
X-ray surveys. The two populations of BL Lacs (radio-selected, RBL, or
X-ray selected, XBL) are known to have significant differences in
parameters, such as the frequencies of the peaks in the Spectral
Energy Distribution (SED)~\citep[e.g,][]{Padovani1995ApJ...444..567P}
and optical fractional polarisation,
$p$~\citep[e.g,][]{Jannuzi1994ApJ...428..130J}. Compared to XBLs, the
RBLs show more prominent radio cores, stronger polarisation and flux
variability, as well as relatively higher luminosities in the radio
and optical. It is presently unknown if there exists a subset of BL
Lacs which is radio-quiet, by analogy to the abundant population of
radio-quiet quasars.~\citet{Stocke1990ApJ...348..141S} noted that
unlike the radio dichotomy of quasars, there is no evidence for
populations of BL Lac objects distinguished by radio
loudness. Contrary to the traditional radio and X-ray surveys for BL
Lac objects, optical surveys for radio-quiet weak-line quasars
(RQWLQs) can, in principle, provide a more effective check on the
existence of radio-quiet BL Lac objects, besides yielding a more
complete census of the BL Lac population as a whole. The existence of
such radio-quiet BL Lacs, if any, could also provide powerful
constraints on the models of jet formation in AGN.\par The Sloan
Digital Sky Survey~\citep[SDSS;][]{Collinge2005AJ....129.2542C} and
the Two Degree Field QSO Redshift
Survey~\citep[2QZ;][]{Londish2002MNRAS.334..941L} resulted in the
first optically selected samples of BL Lac candidates, with 386
candidates coming from the SDSS and 56 from the 2QZ. Since
then,~\citet{Plotkin2010AJ....139..390P} have derived a larger sample
of 723 optically selected BL Lac candidates using the SDSS Data
Release 7~\citep[DR-7,][]{Abazajian2009ApJS..182..543A}. Majority of
the objects in these samples of BL Lac candidates have weak broad
emission-lines with a rest-frame EW $<15.4$~\AA~for the
Ly$\alpha$$+$N{\sc v} emission-line complex
~\citep{DiamondStanic2009ApJ...699..782D} and $<11$~\AA~, $<4.8$~\AA~
for the Mg~{\sc ii} and C~{\sc iv},
respectively~\citep{Meusinger2014A&A...568A.114M}. These have been
termed as weak line quasars (WLQs), however the physical cause of
their abnormally weak emission lines is still debated~\citep[see,
  e.g.,][]{Nicastro2003ApJ...589L..13N,Stalin2005MNRAS.359.1022S,DiamondStanic2009ApJ...699..782D,
  Elitzur2009ApJ...701L..91E,Hryniewicz2010MNRAS.404.2028H,
  Plotkin2010ApJ...721..562P,Laor2011MNRAS.417..681L,
  Liu2011ApJ...728L..44L,Nikolajuk2012MNRAS.420.2518N}. One possible
explanation for the abnormality is that the covering factor of the
broad-line region (BLR) in WLQs could be at least an
order-of-magnitude smaller compared to that in normal
QSOs~\citep[e.g.][]{Nikolajuk2012MNRAS.420.2518N}. An extreme version
of this scenario is that in WLQs the accretion disk is relatively
recently established and hence a significant BLR is yet to
develop~\citep{Hryniewicz2010MNRAS.404.2028H, Liu2011ApJ...728L..44L}.
A yet another possibility is a high mass of the central BH (M$_{BH} >~
3\times10^9 M_{\odot}$) which can result in an accretion disk which is
too cold to emit sufficient number of UV ionizing photons, even when
its optical output is high (\citealt{Laor2011MNRAS.417..681L}, also,
\citealt{Plotkin2010AJ....139..390P}).\par

While the above mentioned limited empirical evidence and theoretical
scenarios are consistent with the bulk of the RQWLQs population being
a special case of the standard radio-quiet quasars, they do not rule
out the possibility that a small subset of this population may in fact
be the long-sought radio-quiet BL Lac objects in which optical
emission does arise predominantly from a relativistic jet of
synchrotron
radiation~\citep[e.g,][]{Stocke1981ApJ...245..375S,Stalin2005MNRAS.359.1022S,
  DiamondStanic2009ApJ...699..782D,Plotkin2010AJ....139..390P}. One
strategy to pursue such a search is to characterize the optical
polarisation and intra-night optical variability (INOV) of
RQWLQs. These are the two main characteristic properties of BL Lac
objects; namely an optical polarisation, $p$ $>$
3\%~\citep[e.g.,][]{Heidt2011A&A...529A.162H} and an INOV duty cycle
approaching the level of $\sim$50 percent (at an INOV amplitude $\psi
>3$\%)~\citep[e.g,][]{GopalKrishna2003ApJ...586L..25G,Sagar2004MNRAS.348..176S,2004MNRAS.350..175S,Stalin2004JApA...25....1S,GopalKrishna2011MNRAS.416..101G,Goyal2012A&A...544A..37G}.
To characterize INOV of RQWLQs, our group has recently analysed a
sample of $70$ intranight light curves of $34$
RQWLQs~\citep{Gopal2013MNRAS.430.1302G,Chand2014MNRAS.441..726C,Kumar2015MNRAS.448.1463K,
  Kumar2016MNRAS.461..666K,Kumar2017MNRAS.471..606K} and reported that
two of the RQWLQs in two separate sessions exhibited strong INOV with
$\psi >$ 10\%, like that of BL Lacs, though the duty cycle of INOV was
found to be only about $\sim 5\%$ which is typical of normal
radio-quiet QSOs (\citealt[][and references therein, see
  also]{Kumar2017MNRAS.471..606K}~\citealt{Liu2015A&A...576A...3L}). On
the other hand, efforts for a systematic polarimetric characterization
of {\it radio-quiet} BL Lac candidates have been quite limited,
although polarimetric surveys have been reported for WLQs in general
~\citep[e.g,][]{Smith2007ApJ...663..118S,
  Heidt2011A&A...529A.162H}. Therefore it seems worthwhile to
investigate a sub-sample of optically selected WLQs whose members have
(i) a radio-loudness parameter{\footnote{Radio-loudness is usually
    parametrised by the ratio (R) of flux densities at 5 GHz and at
    2500\AA~in the rest-frame, and R is $<$ 10 for radio-quiet
    quasars~\citep[e.g.,][]{Kellermann1989AJ.....98.1195K}.}} (R)
$<10$, to ensure radio-quietness and (ii) a secure redshift
measurement, or at least a proper motion consistent with zero. The
proper motion check is particularly relevant for WLQs in view of the
weakness or near absence of features in their spectra, which often
renders their redshift estimates fairly uncertain. For such a sample
of RQWLQs (see Sect. 2), we present here polarimetric and
spectroscopic observations taken with the European Southern
Observatory (ESO) 3.6m telescope at La Silla, Chile. Our main goal is
four-fold: (i) with the new spectroscopic observations of our RQWLQ
sample, we aim to constrain the emission redshifts based on any weak
emission feature(s) that might show up in case the source happens to
be in a low state on account of a weak continuum boosting, (ii) to
investigate spectral properties such as the distribution of spectral
slope, for comparison with the control samples of blazars and normal
QSOs, (iii) to test whether any of the RQWLQs show a strong
polarisation like the classical BL Lacs for which $p > 3-4\%$
typically; this would establish them as a bona-fide radio-quiet
counterpart of BL Lacs, and (iv) to investigate the long-term optical
variability of these RQWLQs, for comparison with blazars and normal
QSOs.\par

This paper is organized as follows. Sect. 2 describes our sample of
RQWLQs and the polarimetric/spectroscopic observations, while Sect. 3
gives details of the data reduction and analysis. In Sect. 4, we
present our results on the spectral and polarisation properties and
the long-term optical variability of the RQWLQs and a comparison with
those already established for blazars and QSOs. A brief discussion of
our results followed by conclusions is presented in Sect. 5.
\begin{table*}
\begin{minipage}{500mm}
{
\caption{The observed sample  (see text).
\label{tab:source_info}}
\begin{tabular}{lcccrc clcc}
\hline
\multicolumn{1}{l}{SDSS Name} & RA (J2000) & DEC(J2000)                       &{$m_{R}$} & PM{\footnote {Proper motion (PM) taken from Gaia-DR2~\citep{Gaia2018arXiv180409365G}. Targets with significant PM\\ are shown in boldface}.} & $z_{emi}${\footnote { Reference is~\citet{Hewett2010MNRAS.405.2302H} for `$\dagger$' and SDSS for $\ddagger$.}}& Obs. Date{\footnote {The year and month for both spectroscopic and
    polarisation observations are the same, but the dates differ\\ (given inside brackets for the polarimetry). Day column marked with two dots (..) refer to no observation.}} & Exp. Time{\footnote {Exposure time for spectroscopic observation.}}\\
         & (h m s)      &($ ^{\circ}$ $ ^{\prime}$ $ ^{\prime\prime }$) & (mag) &(mas/yr)&& Spec.(Pol.) & (sec) \\
 (1)     &(2)             &(3)                             &(4)     &(5)   &(6)   &(7)&(8)\\
\hline
\multicolumn{5}{l}{}\\
J100253.23$-$001727.0  & 10:02:53.23& $-$00:17:27.00 & 19.13      &{\bf 101.46$\pm$0.83}   &                            &2006.04.24(26)&1200x3\\
J102615.30$-$000630.2  & 10:26:15.30& $-$00:06:30.22 & 19.26      &{\bf 52.62$\pm$1.37}  &                            &2006.04.24(27)&1200x3\\
J103607.52$+$015659.0  & 10:36:07.52& $+$01:56:59.05 & 18.80      &0.56$\pm$0.58                      & 1.8768$\pm0.0014^{\dagger}$  &2006.04.24(26)&1200x3\\
J104519.72$+$002614.3  & 10:45:19.72& $+$00:26:14.38 & 18.63    &{\bf 64.81$\pm$0.65}    &                            &2006.04.25(27)&600x3\\
J105355.17$-$005537.7  & 10:53:55.17& $-$00:55:37.72 & 19.41    &{\bf 17.92$\pm$1.12}            &                             &2006.04.25(26)&1500x3\\
J113413.48$+$001042.0  & 11:34:13.48& $+$00:10:42.05 & 18.44      &0.25$\pm$1.07                      & 1.4857$\pm0.0007^{\dagger}$  &2006.04.24(27)&600x3\\
J114554.87$+$001023.9  & 11:45:54.87& $+$00:10:23.96 & 19.54    &{\bf 16.71$\pm$1.14}                        &                            &2006.04.25(27)&1800x3\\ 
J114521.64$-$024757.4  & 11:45:21.64& $-$02:47:57.48 & 18.65      &{\bf 14.85$\pm$0.89}    &                            &2006.04.24(27)&600x3\\ 
J115909.61$-$024534.6  & 11:59:09.61& $-$02:45:34.65 & 19.02    &0.64$\pm$0.83             & 2.0136$\pm0.0006^{\dagger}$  &2006.04.25(26)&1200x3\\
J120558.27$-$004217.8  & 12:05:58.27& $-$00:42:17.82 & 18.75      &{\bf 111.95$\pm$0.84}  &                            &2006.04.24(27)&900x3\\ 
J120801.84$-$004218.3  & 12:08:01.84& $-$00:42:18.31 & 19.05    &{\bf 25.37$\pm$1.02}    &                            &2006.04.25(27)&600x3\\
J122338.05$-$015617.1  & 12:23:38.05& $-$01:56:17.14 & 19.11    &{\bf 28.15$\pm$0.83}     &                            &2006.04.25(..)&1200x3\\
J123437.64$-$012951.9  & 12:34:37.64& $-$01:29:51.94 & 19.18     &0.98$\pm$1.27                       & 1.7105$\pm0.0003^{\ddagger}$ &2006.04.25(27)&1200x3\\ 
J125435.81$-$011822.0  & 12:54:35.81& $-$01:18:22.03 & 19.15     &{\bf 19.97$\pm$0.94}                       &                            &2006.04...(28)& \\
J130009.93$-$022559.2  & 13:00:09.93& $-$02:25:59.27 & 18.90      &{\bf 87.82$\pm$0.70}  &                            &2006.04.24(..)&1200x3\\ 
J140916.33$-$000011.3  & 14:09:16.33& $-$00:00:11.31 & 18.65      &{\bf 38.76$\pm$0.64}  &                            & 2006.04.24(..)&600x3\\
J141046.75$-$023145.3  & 14:10:46.75& $-$02:31:45.35 & 18.13      &{\bf 22.06$\pm$1.66}   &                            &2006.04.24(27)&1200x3\\ 
J142505.59$+$035336.2  & 14:25:05.59& $+$03:53:36.23 & 18.74      &0.11$\pm$0.76                      & 2.2353$\pm0.0012^{\ddagger}$   &2006.04.25(28)&600x3\\ 
J154515.77$+$003235.2  & 15:45:15.77& $+$00:32:35.26 & 18.82      &0.22$\pm$0.70                      & 1.0511$\pm0.0005^{\dagger}$  &2006.04.24(25)&1200x3\\
                       &            &                &           & 			  &                            &2006.04.24(25)&1200x3\\               
                                                                               
\hline
\end{tabular}
}
\end{minipage}
\end{table*}
\section{The Sample and Observations} In Table 1 of their paper, 
\citet{Londish2002MNRAS.334..941L} have given a sample of 56 objects
derived from the 2QZ QSO survey on the criteria of a featureless
optical spectrum and non-significant proper motion.
Likewise,~\citet{Collinge2005AJ....129.2542C} used the SDSS to extract
386 optically selected BL Lac candidates. Out of these total 442 WLQs
we extracted a set of 111 WLQs (88 out of the 386 and 23 out of the 56
candidates) for polarimetry/spectroscopy, by limiting to (i) 8-17h
right ascension range, and (ii) m$_R < 20$-mag. Since our observations
were scheduled for April, we trimmed our list from 111 to 19 RQWLQs,
as listed in Table~\ref{tab:source_info}, by accepting only those
lying within the 10-15h range in right ascension and at sufficiently
low declination for ESO La Silla observatory, and also brighter than
19.5-mag in R-band ($m_{R}$), as well as lacking any published
polarisation measurement (except for two sources). Moreover, the
selected 19 RQWLQs either have a radio-loudness parameter R $<$
10~\citep[see,][] {Kellermann1989AJ.....98.1195K}, or a non-detection
in the FIRST survey (i.e., somewhat conservatively, $<$ 1 mJy at 1.4
GHz, see Becker et al. 1995). Fifteen of these 19 RQWLQs could be
covered in both our polarimetric and spectroscopic observations, 3 in
spectroscopy alone, and one in polarimetry alone (see Table
~\ref{tab:source_info}). These observations were carried out during
24-28 April, 2006 with the ESO 3.6m telescope at La Silla, equipped
with the ESO Faint Object Spectrograph and Camera
~\citep[EFOSC;][]{Buzzoni1984Msngr..38....9B}.\par

Details of the sample of 19 RQWLQs and the observation log are given
in Table~\ref{tab:source_info}. The first four columns list the source
name, right ascension (RA), declination (DEC) and magnitude in R-band
($m_{R}$). Since, for the present purpose we are only interested
  in genuine QSOs we shall subject the sample of 19 RQWLQ candidates
  to additional checks based on the latest proper motion data from
  Gaia measurements:
  DR2~\citep{Gaia2016A&A...595A...1G,Gaia2018arXiv180409365G}.
  This is important since some classes of white dwarfs also display
  essentially feature-less optical spectra, but due to their
  association with the galactic halo population, they are very likely
  to have a finite proper motion. Indeed, 13 out of our 19
  sources are now found to exhibit a non-zero proper motion, based on
  the Gaia/DR2. All 13 are henceforth discounted as RQWLQ
  candidates. The genuineness of the remaining 6 candidates as RQWLQ
  (with zero proper motion) is further confirmed by the availability
  in the literature of accurate, spectroscopically determined
  redshifts (all of which are above z $>$1.0, see
  Table~\ref{tab:source_info}). Our polarimetric and spectroscopic
  observations using the ESO 3.6m telescope are presented below for
  all these 6 RQWLQs.
The last two columns of Table~\ref{tab:source_info} give the
observation dates (for spectroscopy and polarimetry) and the exposure
time (for spectroscopy).\par In addition, for making a statistical
comparison of the optical spectral index and variability behavior of
RQWLQs with those found for blazars and normal QSOs, we have enlarged
the RQWLQ sample by adding the well-defined sample of 34
optically selected RQWLQs which has been derived
from~\citet{Plotkin2010AJ....139..390P} and
~\citet{Meusinger2014A&A...568A.114M} and being followed up under a
separate program to determine their INOV characteristics~\citep[for
  details, see][]{Kumar2015MNRAS.448.1463K}.
\section{Data Reduction And Analysis} 
The polarimetric and spectroscopic observations were carried out
during 24-28 April 2006 at ESO La Silla, Chile, using the 3.6m
telescope equipped with EFOSC~\citep{Buzzoni1984Msngr..38....9B},
under the observation ID 077.B-0822(A) (Table~\ref{tab:source_info}).
The detector system consists of ESO $\#40$ CCD, a Loral/Lesser
Thinned, AR coated, UV flooded, MPP 2048 $\times$ 2048 chip, with a
pixel size of 15 \micron, corresponding to 0.12 arcsec/pixel on the
sky. A chip with a readout noise of $7.8e^{-}$ and a gain of $0.91
e^{-}$/ADU. CCD was used with Bessel -R(\#642) filter. Spectroscopic
observations of $18$ out of our total 19 sources (excluding
J125435.81$-$011822.0) were performed with the EFOSC using Grism 6 and
with a slit width of 1.0 arcsec. Grism 6 provides a wavelength
  coverage from $3860-8070$~\AA~with a wavelength dispersion of
  $2.06$~\AA/pixel and FWHM of around 16.7~\AA~(at 1\arcsec slit-width) resulting
  in spectral resolution of about $\sim$400 (at 6000\AA).

Firstly, all the raw images were corrected for bias subtraction,
flat-fielded and subjected to cosmic-ray removal, using the standard
tasks given in the data reduction software package
IRAF \footnote{\textsc {Image Reduction and Analysis Facility}
  (http://iraf.noao.edu/).}. The raw two-dimensional data were reduced
with the standard procedures using IRAF. The IRAF task $apall$ was
used to extract the spectra. We then carried out the wavelength and
flux calibration using the He-Ar lamp spectra and the standard stars,
respectively. Exposure time for our spectroscopic observations ranges
between 600 to 1800 sec, depending on faintness of the object, as
listed in the last column of Table~\ref{tab:source_info}. We observed
each source in three exposures and finally combined them to improve
the SNR.\par Polarimetric observations for 16 (including the 6 genuine RQWLQs
discussed above) out of our total sample of $19$ sources were carried
out through a Bessel-R filter in the polarimetric mode; i.e., a
Wollaston prism and a half-wave plate were inserted into the beam,
resulting in ordinary and extraordinary images of each target on the
CCD chip, separated by $10$ arcsec. An unpolarised standard star was
also observed for measuring the instrumental polarisation. Seeing
during the observations was close to 1 arcsec. Fluxes of the ordinary
and extraordinary images were measured by aperture photometry using
the IRAF task $phot$, for all four position angles of the Wollaston
prism viz 0$^{\circ}$, 22.5$^{\circ}$, 45$^{\circ}$ and
67.5$^{\circ}$. Stokes parameters (U, Q), percentage polarisation
($p$) and polarisation angle (PA) were then calculated using the
following standard equations:

\begin{equation} 
F(\beta_{i})= \frac{f^{o}(\beta_{i})-f^{e}(\beta_{i})}{f^{o}(\beta_{i})+f^{e}(\beta_{i})}
\end{equation}
where $\beta_{i}$ = 22.5$^{\circ}$ $\times$ ( i - 1) with i = 1, 2, 3, 4 and $f^{o}(\beta_{i})$ and $f^{e}(\beta_{i})$
 are the fluxes measured for the ordinary and the extraordinary images.
The parameters Q, U, $p$ and PA are given by :
\begin{equation} 
Q = \frac{F(\beta_{1}) -F(\beta_{3})}{2}, 
\hspace{0.45cm} U = \frac{F(\beta_{2}) -F(\beta_{4})}{2} 
\end{equation}
\begin{equation} 
p = \sqrt{Q^{2} + U^{2}}
\end{equation}

\begin{equation}
PA = \left\{ \,
\begin{IEEEeqnarraybox}[][c]{ls}
\IEEEstrut
\frac{1}{2}tan^{-1}(\frac{U}{Q}) &    U  $\textgreater$  0,  Q $\textgreater  0$\\
$90$^{\circ}$+$ \frac{1}{2}tan^{-1}(\frac{U}{Q}) & U $\textless$  0,   Q $\textgreater  0$ \\
$90$^{\circ}$+$ \frac{1}{2}tan^{-1}(\frac{U}{Q}) & U $\textless$  0,   Q  $\textless  0$ \\
$180$^{\circ}$+$ \frac{1}{2}tan^{-1}(\frac{U}{Q}) & U $\textgreater$ 0, Q $ \textless 0$
\IEEEstrut
\end{IEEEeqnarraybox}
\right.
\label{eq:example_left_right1}
\end{equation}
\section{Results}
\subsection{Spectroscopic Properties of RQWLQs}
\begin{figure*}
\epsfig{figure=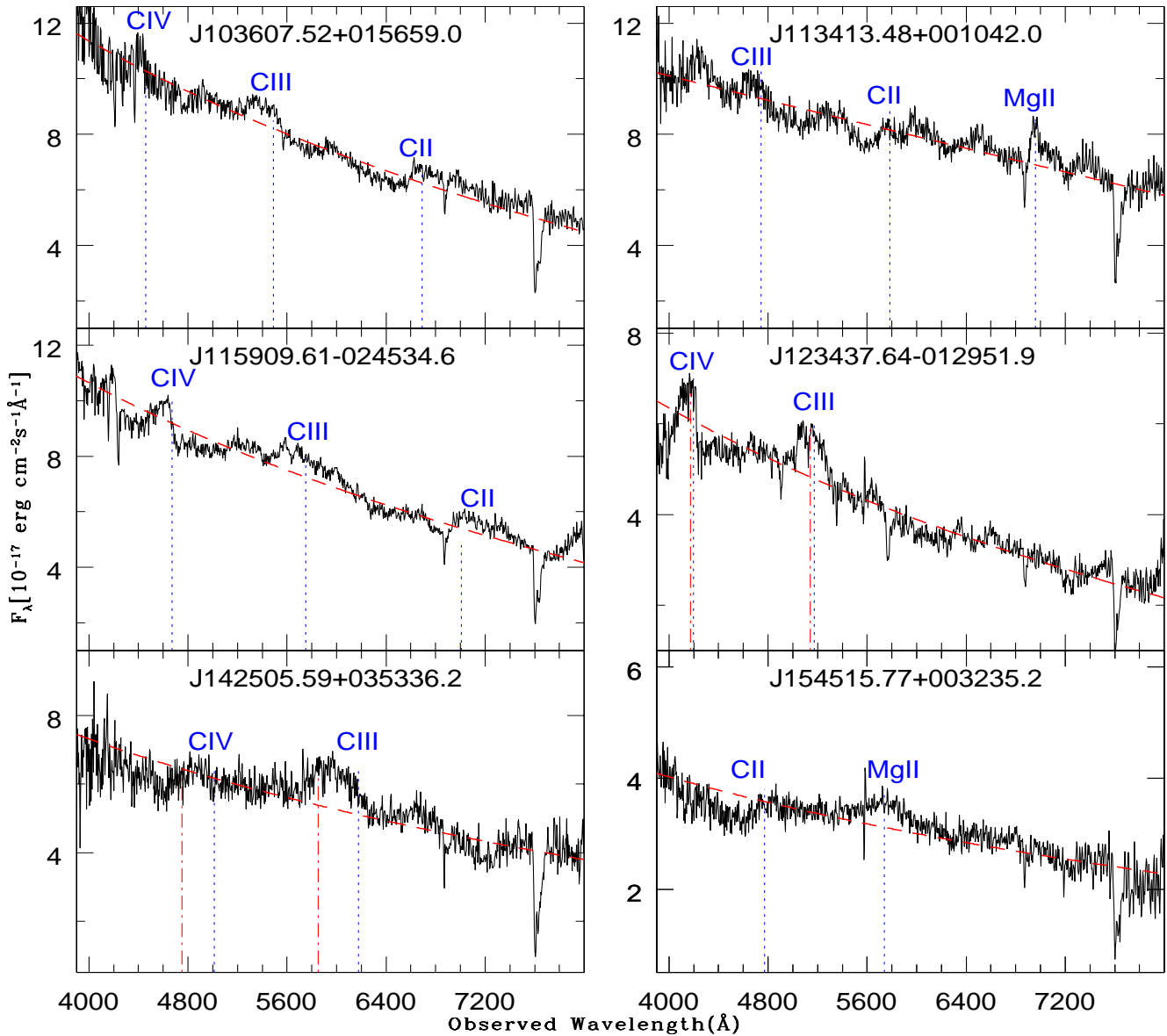,height=16cm,width=18cm,angle=0}
\caption[]{Spectra of our six sources having either zero or insignificant
  proper motion. The present ESO spectra are shown in solid line
  (black), while the spectral fit by a dashed line (red).The small
  dashed lines (blue) mark the positions of the expected emission lines
  based on the SDSS redshifts, while long dashed lines (red) correspond to
  the new redshifts based on our ESO spectra. We note that absorption
  features around 6900\AA~and 7600\AA~are tellric absorption features
  due to the earth's atmosphere.}
\label{fig_spec_no_pm}
\end{figure*}

The motivation behind the present sensitive spectroscopic observations
was not only to measure the spectral shape but also try to
determine/constrain redshift by detecting some weak spectral features
(either absorption or emission), particularly if the object happened
to be in a faint state when the optical continuum is not strongly
Doppler boosted. The reduced spectra for our $6$ RQWLQs are
shown in Fig.~\ref{fig_spec_no_pm}. The spectra of the remaining
  12 sources in our sample, whose extra-galactic origin now seems very
  unlikely in view of the significant proper motion, revealed recently
  by Gaia/DR2, will be presented elsewhere.
The redshifts of all the 6 genuine RQWLQs listed in
Table~\ref{tab:source_info} are based on their SDSS spectra. To assess
the reliability of these existing redshifts, we have marked in
Fig.~\ref{fig_spec_no_pm} the positions of the expected centroid of
various emission lines on our ESO 3.6m spectra. As can be seen from
this figure, the features in our ESO 3.6m spectra are consistent with
the SDSS redshifts for these 6 RQWLQs. We note here that the typical
accuracy of the redshift determination will depend on the spectral
resolution as well as the SNR. The resolution of the SDSS spectra
is about 2000, which is almost 5 times higher than the
resolution of our 3.6m ESO spectra, though in the latter case SNR is
much better. Furthermore, we have noticed that out of our 6 RQWLQs,
the SDSS redshifts of 4 sources have already been improved
by~\citet{Hewett2010MNRAS.405.2302H}. For these sources redshift
estimation could not be improved further using our ESO spectra. The
SDSS redshift of the remaining two sources viz. J123437-012951,
J142505+035336 (which were not covered in the
\citet{Hewett2010MNRAS.405.2302H} redshifts determination), have been
constrained here using our ESO spectra. For instance, our higher SNR
ESO spectrum of J123437-012951 revealed clear signatures of two
emission lines close to $\sim$4200\AA~and $\sim$5200\AA. We use the
publicly available template of composite SDSS spectra (after
appropriate convolution and re-binning to match with our lower
resolution ESO spectra) to determine the redshift of this source based
on the template shifting procedure described
in~\citet{Bolton2012AJ....144..144B}. As can be seen from the middle
right panel of Fig.~\ref{fig_spec_no_pm}, our best-fit redshift of
1.6948$\pm$0.009 is very close to the redshift measured from the SDSS
spectrum of this source. However, for J142505+035336 our new redshift
measurement of 2.0664$\pm$0.0017 using our 3.6m ESO spectrum (based on
the above template fitting technique) is found to be significantly
different from its SDSS redshift (i.e. 2.2353$\pm$0.0012). In
summary, our higher SNR spectroscopic observations (though with a
smaller spectral resolution) have allowed us (i) to reconfirm the
spectroscopic redshift measurement (based on low SNR but a higher
spectral resolution) for 4 of our RQWLQs, (ii) to achieve a moderate
improvement in redshift for the remaining two RQWLQs, besides proving
useful for computing their best-fit power-law spectral slopes.

For fitting the spectral slope, we employed the Levenberg-Marquardt
least-squares minimization technique~\citep[the MPFIT
  package{\footnote{To carry out the simultaneous fitting we have used the
      \textsc{MPFIT} package for nonlinear fitting, written in
      \textsc{Interactive Data Language} routines. MPFIT is kindly
      provided by Craig B. Markwardt and is available at
      http://cow.physics.wisc.edu/\~{}craigm/idl/.}},][]{Markwardt2009ASPC..411..251M}.
In this fitting, we have fitted  a power-law function of the form
$\lambda^{\alpha}$, to describe the AGN continuum emision.
Fig.~\ref{fig_spec_no_pm} shows our spectral fits for the $6$ RQWLQs.

The spectral fitting results enable a statistical comparison of the
spectral slope distribution found for the RQWLQs with those for the
control samples of normal QSOs and blazars. To enlarge the RQWLQ
  sample for this purpose, we have included another 34 well defined
  RQWLQs from~\citet{Kumar2015MNRAS.448.1463K} for which SDSS spectra
  are also available. We also noted that SDSS spectra are available
  also for the 6 RQWLQs of the present sample. So for the sake of
  homogeneous analysis, we have used SDSS spectra for computing the
  spectral slopes of all these 40 RQWLQs.
To build a control sample of normal QSOs, we selected 20 QSOs from
SDSS~\citep{York2000AJ....120.1579Y} for each of the $40$ RQWLQs,
matched in redshift to within $\mid\Delta z\mid$ $<$ 0.005 and in
r-band magnitude to within $\mid\Delta m\mid$ $<$ 0.1, together
resulting in a control sample of $800$ QSOs. Unlike the abundant QSO
population, large catalogs of blazars are not available and hence for
building a control sample of blazars, we have taken all 120 blazars
from the Catalina Real-Time Transient
Survey~\citep[CRTS,\footnote{Catalina Real-Time Transient Survey
    (CRTS), {http://nesssi.cacr.caltech.edu/DataRelease. CRTS
        covers up to $\sim$2500 deg$^{2}$ per night, with 4 exposures
        per visit, separated by 10 min, over 21 nights per
        lunation. All data are automatically processed in real-time,
        and optical transients are immediately distributed using a
        variety of electronic mechanisms. The data are broadly
        calibrated to Johnson V~\citep[see.,][for details]{Drake_crts2009ApJ...696..870D}.}}][]{Drake_crts2009ApJ...696..870D}. This
blazar control sample has an added advantage that it can also be used
for making a comparison of the temporal variability of RQWLQs versus
blazars (Sect. 4.3). For each source in both the control samples
(i.e., of QSOs and blazars) the AGN continuum in the SDSS spectrum,
was fitted with a combination of a power$-$law function and the
Fe~{\sc ii} template, as discussed above for our sample of $40$
RQWLQs.
\begin{figure}
\epsfig{figure=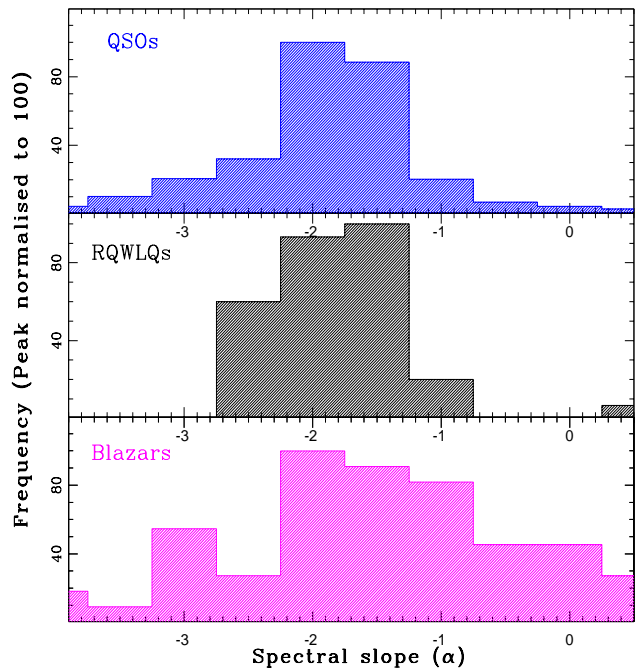,height=9.0cm,width=8.5cm,angle=0}
\caption[]{Spectral slope ($F_{\lambda} \propto \lambda^{\alpha}$) distribution for QSOs (upper panel), RQWLQs (middle panel) and blazars (lower panel).}
\label{fig:spectral_slope}
\end{figure}

\begin{figure*}
\epsfig{figure=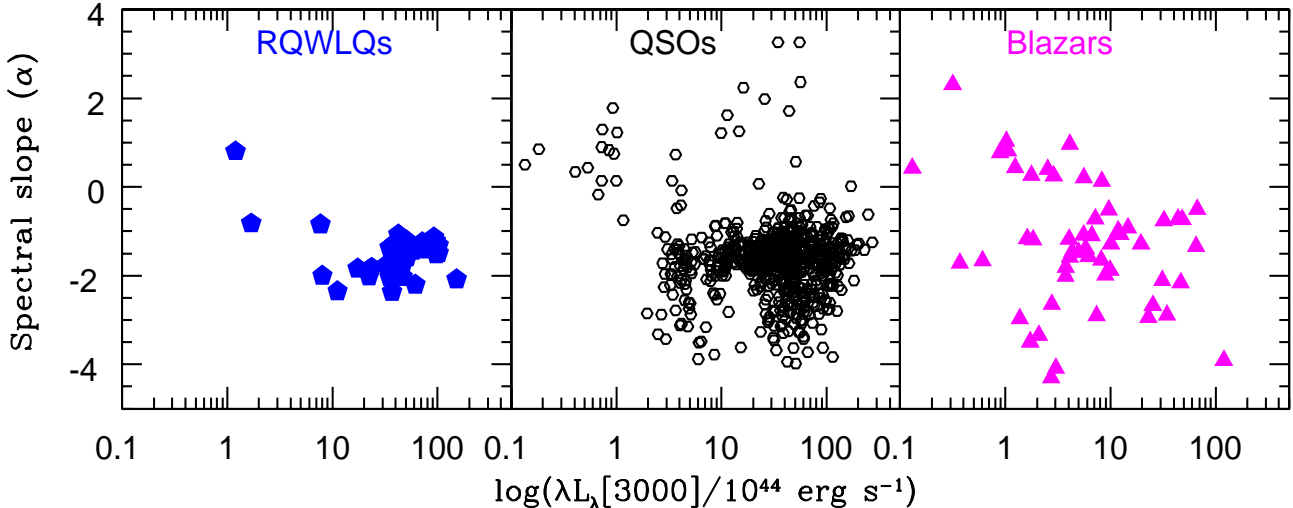,height=19.0cm,width=18.5cm,angle=0}
\vspace{-11.9cm}
\caption[]{Spectral slope ($F_{\lambda} \propto \lambda^{\alpha}$) versus optical luminosity (in the rest-frame) 
for RQWLQs (left panel), QSOs (middle panel)
and blazars (right panel).}
\label{fig:lum_slope}
\end{figure*}
The spectral slope distributions for the 3 AGN classes, viz. RQWLQs,
QSOs and blazars, are shown in Fig.~\ref{fig:spectral_slope}. The
KS-test confirms that RQWLQs and blazars differ at a high level of
significance (97.42$\%$) and so also the QSOs and blazars
(99.97$\%$). In contrast, the difference between RQWLQs and QSOs has a
much lower significance (61.17$\%$).\par Here, we may point out the
possibility of some bias creeping in while comparing the RQWLQs and
blazars, on account of their occupying different luminosity ranges
(since the redshifts of the RQWLQs are usually higher). To examine
this, we show in Fig.~\ref{fig:lum_slope} plots of the spectral slope
versus optical luminosity ($\lambda$L$_{\lambda}$[3000\AA]) for these
three AGN classes. Since no significant correlation is seen between
spectral slope and luminosity, any luminosity bias may not have a
significant influence and the above inferred difference between the
spectral slope distributions for RQWLQs and blazars is likely to be
real (Fig.~\ref{fig:spectral_slope}), suggesting a significant
difference between their emission mechanisms and reinforcing the
premise that RQWLQs are generally closer to QSOs in terms of
emission mechanism.
\subsection{Optical polarisation} 
The results of our polarimetry (Sect. 3) of the 6 bona-fide RQWLQs are
summarized in Table~\ref{tab:polarization_info}.  The first column of
Table~\ref{tab:polarization_info} gives the source name, while the
next 3 columns give the percentage polarisation ($p$), polarisation
angle (PA) and the date of observations. Two of the $6$ sources,
namely J142505.59$+$035336.2 (on 2006.04.28) and J154515.77$+$003235.2
(in the sessions on 2006.04.25 and 2006.04.27) showed polarisation $p$
$> 1\%$, albeit the excess was marginal. The highest observed
polarisation in the present study is $1.59\pm0.53\%$ for
J154515.77$+$003235.2 on 2006.04.27. The two sources have also been
covered in earlier polarimetric studies aimed at identifying BL Lacs
among
QSOs~\citep{Smith2007ApJ...663..118S,Heidt2011A&A...529A.162H}. The
polarisation values reported by these authors for
J142505.59$+$035336.2 and J154515.77$+$003235.2, respectively, are
$<$0.9\%, $<$0.6\% ~\citep{Smith2007ApJ...663..118S} and
$1.1\pm0.6\%$, $<$5.6\% ~\citep{Heidt2011A&A...529A.162H}. These are
consistent with our measurements, suggesting a very mild polarisation
variability for these AGN on year-like time scale.  Both this and the
observed polarisation of around $1\%$ for our RQWLQs is quite low
compared to the average polarisation of around 7\% found for classical
BL Lacs~\citep{Heidt2011A&A...529A.162H}.\par
\begin{figure}
\epsfig{figure=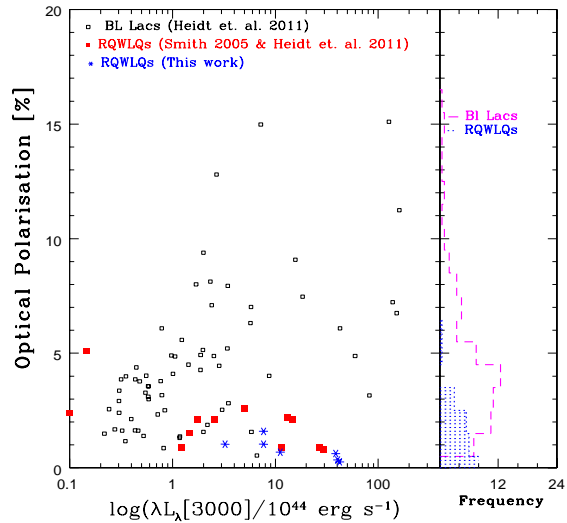,height=8.0cm,width=8.5cm,angle=0}
\caption[]{Optical luminosity versus optical
  polarisation ($p$) of RQWLQs from our sample (blue star),
  from~\citet{Smith2007ApJ...663..118S,Heidt2011A&A...529A.162H}
  (filled square) and the BL Lacs (open square) sample taken
  from~\citet{Heidt2011A&A...529A.162H}.}
\label{fig:lum_pol}
\end{figure}

 Once again, to check the
possible role of bias arising from the difference in optical
luminosity between our RQWLQs and the comparison sample of BL Lacs, we
plot in Fig.~\ref{fig:lum_pol} the polarisation against optical
luminosity for the RQWLQs and BL Lacs. For our $6$ RQWLQs the
polarisation measurement are taken from the present work, supplemented
with the data from ~\citet{Smith2007ApJ...663..118S,
  Heidt2011A&A...529A.162H} for another $13$ RQWLQs which showed
insignificant proper motion~\citep[e.g., see
  also][]{Londish2002MNRAS.334..941L,Collinge2005AJ....129.2542C}. For
the BL Lac comparison sample, we have taken the polarisation data from
the work of ~\citet{Heidt2011A&A...529A.162H}. From
Fig.~\ref{fig:lum_pol}, no correlation is evident between the
percentage polarisation and optical luminosity. The linear Pearson
correlation coefficients are $-0.58$ and $0.44$, respectively, for the
RQWLQs and BL Lacs, which makes it unlikely that the inferred
difference between their polarimetric properties could be an
  artefact of the luminosity difference between the two samples.
\begin{table}
\centering
{
\caption{ The measured fractional polarisation ($p$) and polarisation 
angle (PA) for the 6 RQWLQs. 
\label{tab:polarization_info}}
\begin{tabular}{lcrr}
\hline
\multicolumn{1}{l}{SDSS Name}& $p$ &P.A & Obs. Date\\
           &(\%) &(degree)& (yyyy.mm.dd)  \\
 (1)     &(2)             &(3) & (4) \\
\hline
\multicolumn{4}{l}{}\\
J103607.52$+$015659.0          &0.34$\pm$0.23& \bf{113} & 2006.04.26 \\
J113413.48$+$001042.0          &0.26$\pm$0.28& \bf{168} & 2006.04.27 \\
J115909.61$-$024534.6          &0.62$\pm$0.27& \bf{114} & 2006.04.26 \\
J123437.64$-$012951.9          &0.68$\pm$0.36& \bf{85}  & 2006.04.27 \\
J142505.59$+$035336.2          &1.03$\pm$0.36& \bf{141} & 2006.04.28 \\
J154515.77$+$003235.2          &1.03$\pm$0.61& \bf{35}  & 2006.04.25 \\ 
                               &1.59$\pm$0.53& \bf{66}  & 2006.04.27 \\                                                                               
\hline
\end{tabular}
}
\end{table}
\subsection{Long-term Variability Time Scale}
Structure-function (SF) derived from a light curve is frequently used to infer the variability 
properties, such as the characteristics time-scales and any periodic behavior.
Several definitions of SF have been used in the literature
~\citep{Graham2014MNRAS.439..703G}.
Here we adopt the definition of SF from ~\citet{Hawkins2002MNRAS.329...76H,Schmidt2010ApJ...714.1194S}, and
take an average over all the $N_{lc}$ light curves in a given sample (one light-curve for each object in the sample) as:
\begin{equation}
\begin{tiny}
\label{Eq:sf_info}
S(\tau)= 
\begin{IEEEeqnarraybox}[][c]{ls}
\IEEEstrut
\frac{1}{\sum_{k=1}^{N_{lc}} n^k(\tau)} \sum_{k=1}^{N_{lc}} \left( \sum_{i=1 (i<j)}^{n^k(\tau)} \sqrt{\frac {\pi}{2}}[m^k(t_{j})-m^k(t_{i})] -\sqrt{\sigma_{i}^{2}+\sigma_{j}^{2}}\right)
\IEEEstrut
\end{IEEEeqnarraybox}
\end{tiny}
\end{equation}

where $m^k(t_{i})$ is the magnitude of the $k^{th}$ object at time
$t_{i}$, and the sum runs over the $n^k$($\tau$) time intervals
resulting from all possible combinations of $t_{j}-t_{i}=\tau$, with
j$>$i in the light curve of the $k^{th}$ object in the sample. The
$\sigma_{i}$ and $\sigma_{j}$ are the photometric errors on the
measurements.

\begin{figure}
\epsfig{figure=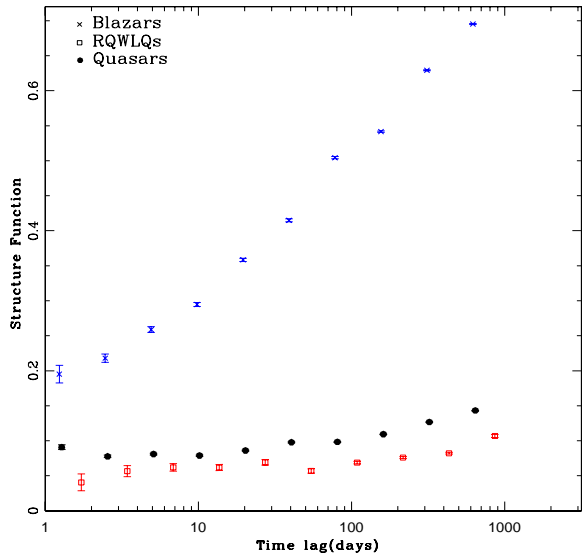,height=8cm,width=8.5cm,angle=0}
\caption[]{Structure function (SF) for RQWLQs (open square), QSOs (filled 
circle) and blazars (blue stars). SFs for RQWLQs and the QSOs are similar, but its 
shape for blazars is very different (see text).}
\label{fig:sf_info}
\end{figure}

Our SF analysis (as per Eq.~\ref{Eq:sf_info}) makes use of the sample
of the $40$ RQWLQs and the corresponding control samples of $800$ QSOs
and $120$ blazars (see Sect. 4.1). The light curves of the RQWLQs,
QSOs and blazars were taken from the CRTS archives, where for each
night we have averaged the available V-band photometric data points
(typically 3), in order to improve the SNR.  The computed SFs for the
3 samples are shown in the Fig.~\ref{fig:sf_info}. Clearly, the SF of
RQWLQs is far better matched to that of the normal QSOs, as compared
to the SF for blazars. This supports the premise that, as a class,
RQWLQs are closer to normal QSOs in term of variability mechanism and
are clearly different from blazars. The K-S test implies that the SFs
for the RQWLQs and blazars differ at a high level of significance
(99.99$\%$), which is also the case when the SFs of the QSOs and
blazars are compared. In contrast, the SFs of RQWLQs and normal QSOs 
are practically indistinguishable.
\section{Discussion and Conclusions}
Polarisation measurements have often been used to validate the BL Lac
classification of AGN. For instance, as mentioned in Sect. 1, optical
polarimetry of a large sample of 182 optically (spectroscopically)
selected BL Lac candidates (both radio loud and radio quiet),
by~\citet{Heidt2011A&A...529A.162H}, showed nearly half of them to be
highly polarised ($p$$>$ 4\%) and all these are radio loud. For
another sample of similarly selected 42 BL Lac
candidates,~\citet{Smith2007ApJ...663..118S} had earlier found that
the sources with undetermined redshift were strongly polarised ($p$
going up to 23\%), whereas at $z > 1$ they found no source having $p >
3$\%. Again, their polarimetric search did not reveal a single good
case of radio-quiet BL Lac.\par In this study we have attempted to
check if any of our 6 candidates, all of which are radio quiet, turns
out to be a bona-fide BL Lac. The polarisation measurements for
  another $13$ of our objects having non-zero proper motions favoring
  a galactic classification will be presented elsewhere. The sensitive
  spectroscopic observations (though at lower spectral resolution), as
  reported here, have allowed us (i) to confirm the existing
  spectroscopic redshifts based on the SDSS spectra for
  4 of our RQWLQs and (ii) to achieve a moderate improvement in
  redshift for the remaining two RQWLQs viz. J123437-012951,
  J142505+035336, without altering the inference on their
  extra-galactic nature consistent with their non-significant proper
  motions. Furthermore, with the SDSS spectra of the 40 RQWLQs
  (including the above 6 RQWLQs) have yielded the best-fit power-law
slopes for their continuum spectra (Sect. 4.1;
Fig.~\ref{fig_spec_no_pm}). This enabled a comparison with the
spectral slope distribution we have determined for the two comparison
samples, consisting of 800 QSOs and 120 blazars, using their SDSS
spectra (Sect. 4.1). This statistical comparison, too, shows that in
terms of optical spectral slope, RQWLQs are much more similar to QSOs
than to blazars (e.g, see Fig.~\ref{fig:spectral_slope} and
Sect. 4.1).\par

In the polarimetric data reported here for the $6$ bona-fide RQWLQs,
only two sources, namely J142505.59$+$035336.2 and
J154515.77$+$003235.2, observed in 3 sessions, were found to have
polarisation $p$ in excess of $1\%$. However, the excess is marginal,
with the highest value being $1.59\pm0.53\%$, measured for
J154515.77$+$003235.2 on 2006.04.27
(Table~\ref{tab:polarization_info}). Polarimetry of both these sources
has earlier been reported by
~\citet{Smith2007ApJ...663..118S,Heidt2011A&A...529A.162H} and their
results are fully consistent with the present measurements, suggesting
that any polarisation variability must be modest.

We have argued that the present comparison of RQWLQs and BL Lacs is
unlikely to suffer from the luminosity bias arising from a majority of
our RQWLQs being located at higher-z and having systematically higher
luminosities compared to the blazar comparison sample. This is because
we do not find a significant dependence of $p$ on optical luminosity
(Fig.~\ref{fig:lum_pol}), both for the blazar and RQWLQ
samples. Therefore, the finding of much stronger optical polarisation
for blazars, as compared to RQWLQs (for which $p$ is found here to be
$<$ $\sim$ 1\%), is likely to be an intrinsic difference. As mentioned
above, polarisation for blazars is generally much higher. Among the 37
classical BL Lac objects observed
by~\citet{Jannuzi1994ApJ...428..130J}, $29$ were found to have $p >
3\%$. The strong contrast again suggests that the mechanism
responsible for the polarisation of the blazar emission plays a minor
role in case of RQWLQs, although polarimetry of large samples of
RQWLQs may still throw up some exceptions, which would be extremely
interesting.\par

Lastly, we have also compared the long-term optical broad-band
(V-band) variability of the 40 RQWLQs with that of the comparison
samples of 800 normal QSOs and 120 blazars, taking the light curves
from the archival CRTS database (Sect. 4.3). A comparison of their
structure functions (Fig.~\ref{fig:sf_info}) provides further support
to the inference that, on the whole, the optical variability mechanism
of RQWLQs is similar to that of QSOs and unlike that operating in
blazars (which are generally more variable at all time-scales from
$\sim$1 to $\sim$1000 days). \par

In summary, from the present study of RQWLQs, employing optical
spectra, polarisation and temporal flux variation on medium to long
time scale, it is evident that the mechanism of RQWLQs central engines
is more similar to that operating in normal QSOs, as compared to
blazars (where the emission is dominated by a Doppler boosted
relativistic jet). A corollary of this would be that the abnormally
weak emission lines in RQWLQs are probably better understood in terms
of the model which invokes a less developed broad-line region with a
low covering factor (Sect. 1). Such a low covering factor of the BLR
in RQWLQs would have an additional consequence, namely an enhanced
evaporation of dust in the torus, leading to a diminished infrared
output. This would be in accord with the up-to 30-40$\%$ lower
infrared emission detected by~\citet{DiamondStanic2009ApJ...699..782D}
in their observations of two RQWLQs. A confirmation of this using a
larger sample of RQWLQs would be very desirable.
\section*{Acknowledgments}
We thank the referee Prof. Michael Strauss for his critical comments
and helpful suggestions on the manuscripts. We gratefully acknowledge
the help from the staff of European Southern Observatory (ESO), in our
observations made under Program ID No. 077.B-0822(A) using the EFOSC
on the 3.6 m telescope operated at the La Silla Observatory. Funding
for the SDSS and SDSS-II has been provided by the Alfred P. Sloan
Foundation, the Participating Institutions, the National Science
Foundation, the U.S. Department of Energy, the National Aeronautics
and Space Administration, the Japanese Monbukagakusho, the Max Planck
Society, and the Higher Education Funding Council for England. The
SDSS Web Site is http://www.sdss.org/.  The SDSS is managed by the
Astrophysical Research Consortium for the Participating
Institutions. The Participating Institutions are the American Museum
of Natural History, Astrophysical Institute Potsdam, University of
Basel, University of Cambridge, Case Western Reserve University,
University of Chicago, Drexel University, Fermilab, the Institute for
Advanced Study, the Japan Participation Group, Johns Hopkins
University, the Joint Institute for Nuclear Astrophysics, the Kavli
Institute for Particle Astrophysics and Cosmology, the Korean
Scientist Group, the Chinese Academy of Sciences (LAMOST), Los Alamos
National Laboratory, the Max-Planck-Institute for Astronomy (MPIA),
the Max-Planck-Institute for Astrophysics (MPA), New Mexico State
University, Ohio State University, University of Pittsburgh,
University of Portsmouth, Princeton University, the United States
Naval Observatory, and the University of Washington.\par This research
has made use of the NASA/IPAC Extragalactic Database (NED) which is
operated by the Jet Propulsion Laboratory, California Institute of
Technology, under contract with the National Aeronautics and Space
Administration. This work has made use of data from the European Space
Agency (ESA) mission {\it Gaia}
(\url{https://www.cosmos.esa.int/gaia}), processed by the {\it Gaia}
Data Processing and Analysis Consortium (DPAC,
\url{https://www.cosmos.esa.int/web/gaia/dpac/consortium}). Funding
for the DPAC has been provided by national institutions, in particular
the institutions participating in the {\it Gaia} Multilateral
Agreement.
\bibliography{references}
\label{lastpage}
\end{document}